\documentclass{article} 
\usepackage{graphicx}
\begin{document}
\title{Quantum nature of time -- proposition of experimental verification}

\author{Slobodan Prvanovi\'c $^1$ and Du\v san Arsenovi\'c $^1$\\
$^{1}$ Institute of Physics Belgrade,\\ National Institute of the Republic of Serbia\\
}

\maketitle
\begin{abstract}
{In quantum mechanics time usually appears as classical parameter which means that it is treated as being essentially different from spatial coordinates that are represented by operators. On the other hand, relativity theory demands to treat space and time on an equal footing and there are many approaches to the problem how to introduce time operator. In order to support these approaches we propose the experiment that will clearly expose the equivalence of space and time regarding the relation towards quantum interference.
The proposed experiment can show that it is possible to have superposition of being created in two different moments in time, and by measuring standard interference pattern, it will be shown that there is interference in time. It is similar to the case of verification of the possibility of single quantum system to be in the state that is superposition of two different positions in space.
If so, this would mean that there should be exactly the same formal approach to the space coordinates and time in quantum mechanics since they are all quantum in their nature.}
\end{abstract}



\section{Introduction}

Time in quantum mechanics is usually taken to be a classical parameter, not an operator, and in this way it differs from the spatial coordinates. The theory of 
relativity, on the other side, demands to treat time and space on an equal footing since they form common four dimensional space-time. If so, space and time have to 
have the same character, or to put it in a different words - they must be of the same nature. Then, the question rises - is the time classical variable or quantum observable, and this is something that should be resolved by some experiment.

What we mean by the quantum nature of time is exactly the same what is meant by the quantum nature of spatial coordinates. The spatial coordinates of quantum 
mechanical system are observables, not variables as they are in the case of classical system. There are observables that are non-compatible with the coordinates, 
so if the time is quantum observable, it should have its own non-compatible observable (which would be the energy).

There are attempts to treat time as quantum observable, {\it i. e.}, there are articles dealing with different operators introduced in order to represent time, 
see references \cite{ref7,ref8,ref9,ref10}. When one is talking about quantum time, one actually presumes that there is possibility to superpose the states that represent different 
moments of time. If we represent the state of the quantum system by statistical operator, in the case of superposition we will have "off-diagonal" elements and, 
as a consequence, there will be a non-commutativity among some operators. None of these is characteristic for classical systems. Single classical system cannot be 
at two different points at one moment of time, superposition of states that represent position of classical system is not possible and there is no non-commutativity 
in classical mechanics.

The problem how to prove the quantum nature of time could be answered, we believe, by performing the slightest possible modification of the standard interference 
experiment. The famous double slit interference experiment is the one that demonstrates that the single quantum system might be at two different places at the 
same moment. By modifying this experiment, we want to propose how to check whether something related to single quantum system can happen at the same point, but 
in different times.

\section{Modification of double slit experiment}

Let us review the standard interference experiment in more descriptive than rigor way and without going in discussion regarding technical details that are not 
important for the present considerations.

As is usually stated, interference pattern on the screen in a double slit experiment proves that involved physical system behaved as a wave on its way towards 
the screen. This means that it passed through both slits simultaneously, {\it i. e.}, it was at two different places at one moment of time. The crucial difference 
between quantum and classical mechanics manifests here in the fact that we can, by lowering intensity of the incident beam, get the interference pattern even in 
the case when just one system is present in the experimental setting. If we consider single system double slit interference experiment, by finding the interference 
pattern after many repetition of the experiment, it is proved that a system, when it was in the region where the plate with the slits is, it was in a superposition 
of states, that quantum system interferes with itself, or that it "goes" through both slits, so to say. 

In order to be able to represent the superposed state, we have to have Hilbert space, {\it i. e.}, phase space of classical mechanics is not suitable mathematical
arena for this kind of considerations. On the other side, the states that are involved in superposition are connected to the position in space. Put in anther way,
the interference pattern of standard interference experiment shows that quantum system can be in superposition of states that represent different points of space,
which means that states representing space points can be superposed. We formalize this by superposition:
$$
\vert q_A \rangle + \vert q_B \rangle .
$$

\begin{figure}[h]	
\includegraphics[width=8 cm]{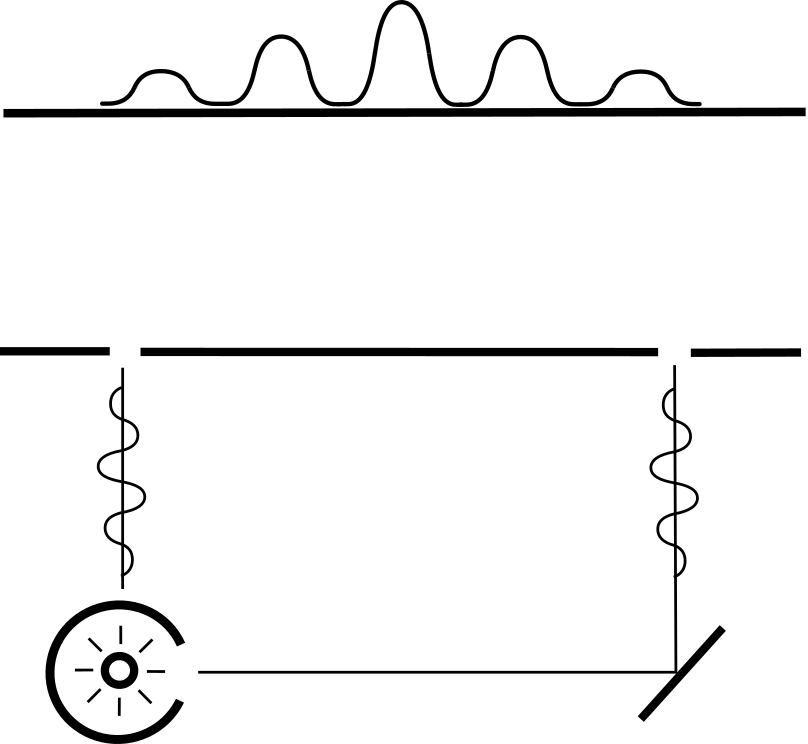}
\centering
\caption{Double-slit experiment with uncertainty in time.\label{fig1}}
\end{figure}  

Of course, it is important not to destroy interference by measuring through which slit the system passed. For the sake of completeness, let us just mention that 
if the system under consideration is initially prepared in a state that is a pulse of a limited duration, then one can expect interference pattern only at the points 
of a screen where pulses coming from two slits overlap. Otherwise, if the pulse from one slit comes at some point after the pulse from the other slit, then 
there will be no interference at that point.

This was the shortest possible review of the most important features of a well known double slit interference experiment. Let us now introduce the slightest 
modifications of this experiment which will transform it from the one that shows spatial interference to the one that addresses possibility of temporal interference.

We can place the source of quantum systems within the container with one hole and let us suppose that this container is rotating, see Fig. 1. Moreover, let us 
suppose that experimental seting is such that there are only two ways for the systems, emitted from the source in the container, to reach the screen. One way 
is to go through the one slit, taking the longer path, and the other is to go through the other slit by taking the shorter path toward the same point at the screen 
where we detect the incoming systems. Let us further suppose that the path difference and rotation of the container are appropriately adjusted. That is, if it 
takes $\Delta T$ to get the container from the position that allows the systems to take the longer path to the position that offers them the shorter one, then the 
path difference should be such that the system emitted at $t_1$, that took the longer path, and the system emitted at $t_2 = t_1 + \Delta T$, that took the shorter 
path, reach the appropriate slits at the same moment.

Finally, if we are dealing with the incident pulses, let us assume that the duration of the pulse is much shorter than $\Delta T$, so if the pulse starts at the 
moment when the opening of the container is oriented towards the longer path, it will certainly end before the opening gets the position that offers the systems 
the other, shorter path towards the screen.

So, the experiment we are proposing and the standard ones differ in the first half of the systems voyage from the source to the screen. In the case of the standard 
ones, the system is emitted at some moment, then it takes both paths as a wave, interferes with itself and ends at some point of the screen. In the present case, 
the story is a bit different. Namely, in order to interfere with itself, the system has to be at the same moment at both slits. Now, this can happen only if the 
system is emitted in the superposition of states that represent the moments $t_1$ and $t_1 + \Delta T$. In other words, two paths are not determined by two slits, 
but by two positions of the container, and this follows from the moments at which the system is emitted. If the system is emitted in the superposition of states 
that represent the moments $t_1$ and $t_1 + \Delta T$, then the longer path, that is related to the moment $t_1$, and the shorter path, that is related to the 
moment $t_1 + \Delta T$, are in superposition even before the system reaches the slits.

If the system is emitted at one of the moments $t_1$ or $t_1 + \Delta T$ only, then it will reach the screen, but there will be no interference.  If it is emitted at 
any other moment, the system will not reach the screen. So, if we get the interference pattern at the end of the experiment (after many repetitions, i.e., after 
enough systems have reached the screen so we can recognize the pattern), then we can conclude that the superposition of states that represent moments in time is 
possible and actually realized since there is only one way to get the interference pattern at the end of the experiment. And, to have superposition of vectors 
representing moments in time means to deal with the time as a quantum observable, not the classical variable.

\section{Operator of time}

If we find out that there is superposition of states that represent different moments in time, then we can conclude that time is on an equal footing with space 
regarding their quantum structure. That is, we can conclude that we have to introduce states $\vert t \rangle$, operator of time $\hat t$, the conjugate operator 
$\hat s$, commutation relations etc. All of these should be introduced in a way that is completely similar with the one used for the spatial coordinates in the 
standard quantum mechanics. The reason is twofold. Firstly, there is no essential difference between spatial and temporal interference and, secondly, according 
to the theory of relativity, space and time should be on an equal footing. Therefore, the state of the quantum system that should be used as formal representation 
of temporal interference is:
\begin{equation}
\vert \psi \rangle = c_A \vert t_A \rangle + c_B \vert t_B \rangle ,
\end{equation}
where, if it should address the proposed experiment, $t_B = t_A + \Delta T$. This means that there is a rigged Hilbert space ${\cal H}_t$
 with vectors $\vert t \rangle$, where $t \in {\bf R}$. Like in the case of spatial degrees of freedom, in ${\cal H}_t$ acts operator of time $\hat t$:
\begin{equation}
\hat t = \int t \vert t \rangle \langle t \vert d t .
\end{equation}
There is the operator $\hat s$ that is conjugate to the operator of time:
\begin{equation}
{1\over i\hbar} [\hat t , \hat s ] = - \hat I .
\end{equation}
This is the operator of energy which, just like the operator of time, has continuous spectrum $\{ -\infty , +\infty \}$. If it is represented in $\vert t \rangle 
$ basis, this operator becomes differential operator $i\hbar {\partial \over \partial t } $. Obviously, the operators of time and energy are completely similar to 
the operators of coordinate and momentum. So, for example, the eigenvectors of the operator of energy in $\vert t \rangle$ representation are $ e^{{1\over i\hbar} 
E\cdot t}$, where $E\in \bf R$.

This is short recapitulation of the formalism of the operator of time that we have introduced in \cite{ref1,ref2,ref3,ref4,ref5,ref6}. Our approach to the operator of time is similar to the one 
given in \cite{ref10}, while the other can be found in \cite{ref7,ref8,ref9} and references therein. In order to complete this short review of our approach, let us consider both space and 
time operators. For the case of one degree of freedom, there are $\hat q \otimes \hat I$, $\hat p \otimes \hat I$, $\hat I  \otimes \hat t$ and $\hat I \otimes 
\hat s$, acting in ${\cal H}_q \otimes {\cal H}_t$, and for these self-adjoint operators the following commutation relations hold:
\begin{equation}
  {1\over i\hbar} [\hat q \otimes \hat I, \hat p \otimes \hat I ] = \hat I \otimes \hat I ,
\end{equation}
\begin{equation}
  {1\over i\hbar} [\hat I \otimes \hat t , \hat I \otimes \hat s ] = - \hat I \otimes \hat I .
\end{equation}
The other commutators vanish. In \cite{ref1} we have shown that the Schr\"odinger equation, that appears as constraint in ${\cal H}_q \otimes {\cal H}_t$, selects the 
states with non-negative energy, due to the bounded from below spectrum of the standard Hamiltonians. Namely, the Hamiltonian $H (\hat q, \hat p )$ and $\hat s$ 
are acting in different Hilbert spaces, but there is subspace of ${\cal H}_q \otimes {\cal H}_t$ where:
\begin{equation}
 \hat s \vert \psi \rangle  = H (\hat q, \hat p ) \vert \psi \rangle  .
\end{equation}
The vectors that satisfy this constraint have non-negative energy for the Hamiltonians that are usually used in quantum mechanics and by taking $\vert q \rangle 
\otimes \vert t \rangle$ representation of (6), one gets: the familiar form:
\begin{equation}
 i \hbar {\partial \over \partial t } \psi (q,t)  = \hat H \psi (q,t)  .
\end{equation}
 With the shorthand notation $\hat H  = H (q , -i\hbar {\partial \over \partial q } )$, which is nothing else but the well known form of the Schr\"odinger equation.

\section{Concluding remarks}

The experiment we have proposed might be seen as theoretical idealization, so that is the reason why we have not discussed it in details, especially regarding the 
parameters. Perhaps the feasible experiment that confirms the quantum nature of time will not be as naive as the proposed one, but it should, we believe, address 
the possibility of superposition of states representing moments in time. That is crucial. What characterizes our proposal is that the result of the experiment would 
hold for the whole ensemble. In this way our proposal differs from the gedanken or realized experiments where some important claims apply only to post selected 
sub-ensembles.

There could be many reasons for the absence of the interference pattern. For example, if we, in any way, observe through which slit the quantum systems are passing, 
then the interference will be destroyed. However, if we get the interference pattern in the proposed experimental situation, then there will be no other explanation 
than that the quantum systems have been created in a state that is superposition of the states $\vert t_A \rangle$ and $\vert t_B \rangle$. If so, the quantum 
character of time would be demonstrated.

In the same manner in which we have concluded that the single quantum system can be at two different places at one moment of time, which we have realized by 
considering the standard spatial double slit experiments, we can conclude that something that is unique can happen to the single quantum system at one place, the source, but 
in two moments. This we can find out by the proposed modification of the standard experiments (with the rotating container), and the mentioned "unique something" 
is the act of creation of the quantum system.

When the quantum character of time is confirmed, then the whole machinery of the quantum formalism should be applied to time, and this we have done in the present 
paper by presenting the short recapitulation of our approach to the operator of time. On the other side, by finding that the time is essentially quantum, some 
experiments that are usually related to the position in space might be translated to the appropriate ones which are related to time. So, after the temporal 
interference, it would be interesting to design the experiment with temporal entanglement.

The operator of energy is the one that is conjugated to the operator time. Just like in the momentum-coordinate case, the eigenvector of energy $\vert E \rangle$ can 
be seen as the superposition of uncountably many eigenvectors of time $\vert t \rangle$:
\begin{equation}
\vert E \rangle = \int _{-\infty} ^{+\infty} \exp ({{1\over i \hbar} E \cdot t}) \vert t \rangle dt .
\end{equation}
Regarding the superposition (or coherent mixture in the language of statistical operators), for which it could be said that it is the most important quantum feature, 
this state is not essentially different from (1), within which we have superposition of two eigenvectors of time. Interesting {\it per se} is the interpretation 
of (1), and it can be quite different from the interpretation of the last expression. Namely, for the state (1) it could be said that it represent situation when the 
quantum system is not created at some particular moment of time. That is the state of the quantum system that was created in a superposition of two moments.

Acknowledgments: {The authors want to thank to Mihailo Rabasovi\' c and Aleksandar Krmpot for constructive debate and advices regarding experimental setup and realization.
The authors acknowledge funding provided by the Institute of Physics Belgrade, through the grant by the Ministry of Education, Science and Technological Development of the Republic of Serbia.}

\end{document}